\documentclass[11pt,a4paper]{article}

\usepackage{bm,amsfonts}
\usepackage{graphicx}
\usepackage{hyperref}

\newcommand{\Z}{\mathbb{Z}}
\newcommand{\C}{\mathbb{C}}

\newcommand{\U}{\mathrm{U}}
\newcommand{\SU}{\mathrm{SU}}

\makeatletter

\@addtoreset{equation}{section}
\makeatother

\date{\today}

\begin{document}

\begin{titlepage}

\renewcommand{\thefootnote}{\fnsymbol{footnote}}

\begin{flushright}
RIKEN-MP-39
\\
\end{flushright}

\vskip5em

\begin{center}
 {\Large {\bf 
 Spinless basis for spin-singlet FQH states
 }}

 \vskip3em

 {\sc Taro Kimura}\footnote{E-mail address: 
 \href{mailto:tkimura@ribf.riken.jp}
 {\tt tkimura@ribf.riken.jp}}

 \vskip2em

{\it Department of Basic Science, University of Tokyo, 
 Tokyo 153-8902, Japan\\ \vskip.2em
 and\\ \vskip.2em
 Mathematical Physics Laboratory, RIKEN Nishina Center, Saitama 351-0198,
 Japan 
}

 \vskip3em

\end{center}

 \vskip2em

\begin{abstract}
We investigate an alternative description of the $\SU(M)$-singlet FQH
 state by using the spinless basis.
The $\SU(M)$-singlet Halperin state is obtained via the $q$-deformation
 of the Laughlin state and its root of unity limit, by applying the
 Yangian Gelfand-Zetlin basis for the spin Calogero-Sutherland model.
The squeezing rule for the $\SU(M)$ state is also investigated in terms of the
 spinless basis.
\end{abstract}

\end{titlepage}

\tableofcontents

\setcounter{footnote}{0}


\section{Introduction}
\label{sec:Intro}

Recently it has been shown the Jack polynomial provides a universal
description of the fractional quantum Hall states (FQH states)
\cite{PhysRevLett.100.246802,PhysRevB.77.184502,PhysRevLett.101.246806},
satisfying the admissible condition \cite{Feigin01012002} which
characterizes the generalized statistics of the FQH states.
This description is understood from the viewpont of the conformal field
theory (CFT): the Jack polynomial for the generic $(k,r)$-admissible
condition is regarded as the conformal block of the CFT with the
extended chiral algebra $\mathrm{WA}_{k-1}(k+1,k+r)$
\cite{Bernevig:2009JPhA,1751-8121-42-44-445209,PhysRevB.82.205307}.
It is also applied to the FQH states in the presence of the internal (spin)
degrees of freedom
\cite{Halperin:1983,PhysRevLett.60.956,PhysRevLett.82.5096,Ardonne:2001NuPhB} by
considering the non-symmetric Jack polynomial \cite{Kasatani01012005,springerlink:10.1007/s00220-007-0341-0,Ardonne:2011,Estienne:2011}.

The Jack polynomial is the eigenfunction of the Laplace-Belrtami
operator, which is regarded as the Hamiltonian of the
Calogero-Sutherland model, up to the gauge transforamtion.
If we consider the spin degrees of freedom, we have to deal with the
spin Laplace-Beltrami operator.
In this case there are degeneracies in its spectrum due to the spin
degeneracy.
Thus it is difficult to apply it naively to numerical applications \cite{PhysRevLett.103.206801,PhysRevB.84.045127}.
On the other hand, it is shown that such a degeneracy is resolved by
utilizing the Yangian symmetry of the spin Calogero-Sutherland model
\cite{Takemura:1996qv,Uglov:1997ia}.
Indeed the orthogonal basis for the spin Calogero-Sutherland model is
constructed by applying such a basis. 
The Yangian Gelfand-Zetlin basis for the spin
Calogero-Sutherland model is realized as the root of unity limit of the
$q$-deformed theory \cite{Uglov:1997ia}.
This means we have an alternative method to describe the
$\SU(M)$-singlet FQH state, i.e. the root of unity limit of the
Macdonald polynomial, which is called the Uglov polynomial
\cite{KuramotoKato200908}, instead of the non-symmetric Jack polynomial.

In this paper we investigate a novel description of the FQH states with
the spin degrees of freedom by utilizing the Yangian Gelfand-Zetlin
basis, which is well investigated in \cite{Uglov:1997ia} for the spin
Calogero-Sutherlnad model.
First the $\SU(M)$-singlet condition is discussed by considering the
generic property of the Lie algebra.
We obtain the Fock condition as well as the standard $\SU(2)$-singlet
Halperin state.
Starting from the Laughlin state, 
we construct the spin-singlet Halperin state through the $q$-deformation
of the $\U(1)$ primary field.
This procedure is based on the Yangian Gelfand-Zetlin basis for the spin
Laplace-Beltrami operator, which is well studied from the viewpoint of
the spin Calogero-Sutherland model.
We then show the relation between the $\SU(M)$-singlet state and a certain
spinless state in terms of the occupation number representation.
The squeezing rule for the $\SU(M)$ state is consistently translated
into the spinless basis.
We finally comment on the underlying CFT for the
$\SU(M)$-singlet state, and its relation to the four dimensional gauge
theory on the orbifold and the root of unity limit of the $q$-deformed CFT.

\section{Spin-singlet FQH states}
\label{sec:singlet}

Before discussing the spinful FQHE states, let us first
introduce the most fundamental example of the FQH state, which is called
the Laughlin state \cite{Laughlin:1983fy},
\begin{equation}
 \Phi_{\rm L} (\{z_i\}) = \prod_{i<j}^N (z_i - z_j)^r .
  \label{Laughlin}
\end{equation}
For convenience we now omit the Gaussian factor, $\exp \left[-\sum_i |z_i|^2
/(4\ell^2)\right]$, with the magnetic length being $\ell^2 = \hbar / (eB)$.
The filling fraction of this state is given by $\nu = 1/r$, and the
power $r$ has to be odd integer due to the anti-symmetricity of the
wavefunction for a fermionic system.
Thus this wavefunction gives rise to the odd denominator series of the
FQH states.

Actually this Laughlin state does not include internal degrees of
freedom.
However we will see the spin-singlet FQH states
are built from this fundamental wavefunction.

\subsection{$\SU(2)$ theory}

A natural generalization of this Laughlin state (\ref{Laughlin}) is the
Halperin state \cite{Halperin:1983}, which accompanies $\SU(2)$ spin
degrees of freedom,
\begin{equation}
 \Phi_{\rm H}(\{z_i,w_i\}) = 
  \prod_{i<j}^{N_\uparrow} (z_i - z_j)^r 
  \prod_{i<j}^{N_{\downarrow}} (w_i - w_j)^r
  \prod_{i,j} (z_i - w_j)^s .
  \label{Halperin}
\end{equation}
Here $z_i$ and $w_i$ stand for posisions of up- and down-spin particles,
respectively.
Since we usually investigate the FQH system in the presence of strong magnetic
fields, it is natural to consider fully spin-polarized states.
This means the spin degrees of freedom are frozen in such a case.
However, when there are pseudo-spin degrees, e.g. valley degeneracy,
multi-layer systems and so on, the spinful FQH state would
provide a good description.

We consider the spin operators for this wavefunction to
discuss the spin-singlet state.
When the spin-raising operator $S_{+,i}$ acts on this wavefunction
(\ref{Halperin}), one of the down-spin particles is changed into the
up-spin state, i.e. $w_i \to z_{N_\uparrow + 1}$.
We often omit the index labeling the particle.
If the state is fermionic, it has to be anti-symmetrized with all the
up-spin particles,
\begin{equation}
 S_+ \Phi_{\rm H}(\{z_i,w_i\})
  = \Phi_{\rm H}(\{z_i,w_i\}) 
  - \sum_{j=1}^{N_\uparrow} \Phi_{\rm H} (z_j \leftrightarrow w_i) .
\end{equation}
Thus the spin-raising operator yields
\begin{equation}
 S_+ = 1 - \sum_{j=1}^{N_\uparrow} e(w_i,z_j) 
  \qquad (\mbox{for fermions}) ,
\end{equation}
where the operator $e(w_i,z_j)$ exchanges $w_i$ and $z_j$.
On the other hand, for a bosonic state, the wavefunction has to be
symmetrized with up-spin particles.
This means the sign factor arising in the spin-raising operator is
modified as
\begin{equation}
 S_+ = 1 + \sum_{j=1}^{N_\uparrow} e(w_i,z_j) 
  \qquad (\mbox{for bosons}) .
\end{equation}
Similarly an up-spin particle is changed into the down-spin state under
the spin-lowring operation, $z_i \to w_{N_\downarrow+1}$.
Therefore the spin-lowering operator is written as
\begin{equation}
 S_- = 1 \mp \sum_{j=1}^{N_\downarrow} e(z_i,w_j) .
\end{equation}
The sign factor takes {\em minus} for fermionic and {\em plus} for
bosonic systems.
The other spin operator is simply given by $S_z = (N_\uparrow -
N_\downarrow)/2$.
Therefore the spin-singlet condition, which is given by
\begin{equation}
 S_+ \Phi_{\rm H}(\{z_i,w_i\}) = 0, \qquad
 S_- \Phi_{\rm H}(\{z_i,w_i\}) = 0, \qquad
 S_z \Phi_{\rm H}(\{z_i,w_i\}) = 0,
 \label{singlet_cond}
\end{equation}
yields $r=s+1$ and $N_\uparrow = N_\downarrow$ for the Halperin state
(\ref{Halperin}).
This is just regarded as the Fock condition \cite{PhysRev.125.164}.
This also implies that $r$ is odd/even for fermions/bosons.

\subsection{$\SU(M)$ theory}

We extend the singlet-condition (\ref{singlet_cond}) to arbitrary
$M$-component systems.
A simple generalization of the Halperin state (\ref{Halperin}) is given
by
\begin{equation}
 \Phi_{\rm H}^M (\{z_i^{(u)}\}_{u=1, \cdots, M})
  = \prod_{u=1}^{M} \prod_{i<j}^{N^{(u)}} (z_i^{(u)}-z_j^{(u)})^r
  \prod_{u<v}^M \prod_{i,j} (z_i^{(u)}-z_j^{(v)})^s .
  \label{higher_Halperin}
\end{equation}
This is an $M$-state wavefunction: $z_i^{(u)}$ stands for a position
of an $i$-th $u$-state particle.

To assign a similar manipulation to this wavefunction, we then need the
raising and lowering operators for $\SU(M)$ group.
According to the general theory of the Lie algebra, the generators can be
split into $H_i$, $E_i$ and $F_i$.
Here $E_i$ and $F_i$ are just regarded as the {\em raising} and {\em lowering}
operators, while Cartan subalgebra consists of $H_i$.
They are simply represented by introducing creation and
annihilation operators for the $u$-th state, $a^{(u)\dag}$ and $a^{(u)}$,
\begin{eqnarray}
 H_u & = & a^{(u)\dag} a^{(u)} - a^{(u+1)\dag} a^{(u+1)}
  \qquad (u = 1, \cdots, M-1), \\
 E_{(u,v)} & = & a^{(u)\dag} a^{(v)}
  \hspace{10em} (u > v), \\
 F_{(u,v)} & = & a^{(u)\dag} a^{(v)}
  \hspace{10em} (u < v). 
\end{eqnarray}
This means $E_{(u,v)}$ and $F_{(u,v)}$ convert $v$-th state into $u$-th state.
For $u > v$ and $u < v$, we call it raising and lowering, respectively.
Remark the numbers of these generators are $M(M-1)/2$ for $E$ and $F$, $M-1$
for $H_i$.
Thus the total dimension of $\SU(M)$ becomes $M^2-1$.
For example, $\SU(2)$ algebras are represented as 
$S_+ = a^\dag_\uparrow a_\downarrow$, 
$S_- = a^\dag_\downarrow a_\uparrow$ and
$S_z = (a^\dag_\uparrow a_\uparrow - a^\dag_\downarrow a_\downarrow)/2$.
The total number of the generators is consistent with $\mathrm{dim}\,
\SU(2)=3$.

We then explicitly show the singlet condition for $\SU(M)$ theory.
The corresponding condition to (\ref{singlet_cond}) is given by
\begin{equation}
 H_u \Phi_{\rm H}^M = 0, \qquad
 E_{(u,v)} \Phi_{\rm H}^M = 0, \qquad
 F_{(u,v)} \Phi_{\rm H}^M = 0.
 \label{singlet_cond_M}
\end{equation}
The first one is simply satisfied when $N^{(1)}=N^{(2)}=\cdots=N^{(M)}$.
As the case of $\SU(2)$, the other operators are represented as
\begin{equation}
 1 \mp \sum_{i=1}^{N^{(u)}} e(z^{(u)}_i,z^{(v)}_j)
  = \left\{
     \begin{array}{cc}
      E_{(u,v)} & (u>v) \\ 
      F_{(u,v)} & (u<v)\\
     \end{array}
    \right. .
\end{equation}
Again the sign factor takes minus for fermionic and plus for bosonic systems.
Thus we have an essentially the same condition for $\SU(M)$ theory as
the $\SU(2)$ theory.
The singlet condition is just given by $r=s+1$ and
$N^{(1)}=N^{(2)}=\cdots=N^{(M)}$.

Remark the filling fraction of the Halperin state
(\ref{higher_Halperin}) satisfying this $\SU(M)$-singlet state is given by
\begin{equation}
 \nu = \frac{M}{M(r-1)+1}.
  \label{fraction_M}
\end{equation}

\section{From Laughlin to Halperin}
\label{sec:one_to_multi}

We show the spin-singlet FQH state can be obtained from a certain
spinless state.
Here we concentrate on the Abelian FQH state, i.e. the Laughlin state.
Its validity for the generic cases is discussed in section
\ref{sec:admissible}.

The method we discuss here is based on the Yangian Gelfand-Zetlin basis for
the spin Calogero-Sutherland model \cite{Uglov:1997ia}, which is also applied
to the gauge theory partition function for the orbifold theory
\cite{Kimura:2011zf,Kimura:2011gq}.
We often use the non-symmetric Jack polynomial to describe the spinful model.
However there are degeneracies in its spectrum due to the internal spin
degrees of freedom.
By resolving the spin degeneracy with the Yangian symmetry, we
obtain the alternative basis for the spinful model, which can be written
in terms of symmetric polynomials. 
We apply this method to the FQH states to obtain an alternative way of
describing the spinful FQH states.

\subsection{The $q$-bosonic field}

We introduce the $q$-deformed FQH state by considering the $q$-bosonic field.
The $q$-boson is quite similar to the standard free boson field,
satisfying the slightly modified commutation relations
\cite{Shiraishi:1995rp,Awata:1995zk},
\begin{equation}
 [a_n, a_m] = n \frac{1-q^{|n|}}{1-t^{|n|}} \delta_{n+m,0}, \qquad
 [a_n, Q] = \frac{1}{r} \delta_{n,0},
 \label{comm_rel}
\end{equation}
where we parametrize $t=q^r$.
The corresponding bosonic field is given by
\begin{equation}
 \varphi(z) = Q - a_0 \log z - i \sum_{n\not=0} \frac{1}{n}
  \frac{1-t^{|n|}}{1-q^{|n|}} a_n z^{-n}.
  \label{q_OPE}
\end{equation}

As the case of the standard CFT, the operator product expansion (OPE) of
this $q$-boson plays an essential role in considering the conformal
block, which is interpreted as the FQH wavefunction.
The singular part of the OPE is given by
\begin{equation}
 \varphi(z) \varphi(w) \sim  - \log
  \left[\frac{(w/z;q)_\infty}{(tw/z;q)_\infty} z^r \right],
\end{equation}
where we use the Pochhammer symbol, $(x;q)_n=\prod_{m=0}^{n-1} (1-x
q^n)$.
As a result, the conformal block of the $q$-primary field
$V(z)=\,:\exp(i\varphi(z)):$ is calculated by utilizing this OPE relation,
\begin{equation}
 \left\langle
  V(z_1)  \cdots V(z_N)
 \right\rangle
 = \prod_{i<j}^N \frac{(z_i/z_j;q)_\infty}{(tz_i/z_j;q)_\infty} z_j^r .
 \label{q_Laughlin}
\end{equation}
Remark this is almost the same as the $q$-Vandermonde determinant, which
is just the weight function of the Macdonald polynomial
\cite{Mac_book}, with the extra contribution of the zero mode.
Actually, as well known, the Laughlin state is interpreted as the
conformal block of the $\U(1)$ primary fields.
Thus, we now regard this conformal block (\ref{q_Laughlin}) as the
$q$-Laughlin state.

\subsection{The root of unity limit}

We can easily show the $q$-deformed state (\ref{q_Laughlin}) goes back
to the standard Laughlin state (\ref{Laughlin}) by taking the limit,
$t=q^r$ and then $q\to 1$.
Note that the Macdonald polynomial is reduced to the Jack polynomial in
this limit.

On the other hand, to implement the Yangian Gelfand-Zetlin basis, we
take another limit of the $q$-state, i.e. the root of unity limit,
\begin{equation}
 q \longrightarrow \omega_M q, \qquad
 t \longrightarrow \omega_M t = \omega_M q^r, \qquad
 q \longrightarrow 1,
 \label{rootofunity}
\end{equation}
where $\omega_M = \exp (2\pi i/M)$ is the $M$-th primitive root of
unity.
This limit is singular because there are inifinite products
appearing in the $q$-state, thus its convergence radius is $|q|=1$.
Therefore we have to deal with it appropriately
\cite{Kimura:2011zf,Kimura:2011gq}.

\begin{figure}[t]
 \begin{center}
  \includegraphics[width=35em]{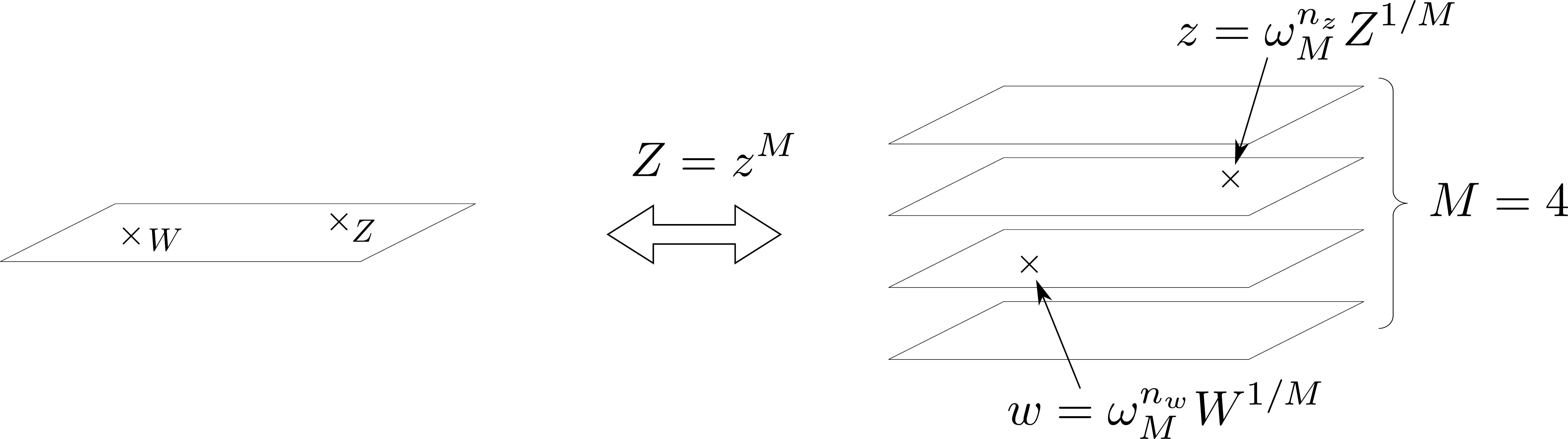}
 \end{center}
 \caption{The branch of $Z=z^M$ for $M=4$. Left and right show
 $Z$ and $z$ planes, respectively.}
 \label{fund}
\end{figure}

Using the formula for the root of unity limit (\ref{rootofunity})
\cite{Kimura:2011zf,Kimura:2011gq},
\begin{equation}
 \frac{(z;q)_\infty}{(tz;q)_\infty} \longrightarrow 
  (1-z^M)^{(r-1)/M}(1-z),
\end{equation}
the OPE of the $q$-state (\ref{q_OPE}) in the limit (\ref{rootofunity})
turns out to be
\begin{equation}
 \varphi(z) \varphi(w) \sim
  - \log (z^M-w^M)^{(r-1)/M}(z-w).
  \label{OPE1}
\end{equation}
To discuss the spin-singlet wavefunction, we now introduce another set of
coordinates, 
\begin{equation}
 z^M = Z, \qquad w^M = W.
  \label{change_variable}
\end{equation}
This means the original one is described as $z = \omega_M^{n_z} Z^{1/M}$,
$w = \omega_M^{n_w} W^{1/M}$, with $n_z, n_w = 0, 1, \cdots, M-1$.
Here $n_z$ and $n_w$ stand for the branches of the $M$-th roots of $Z$
and $W$, respectively, as shown in Fig.~\ref{fund}.
We rewrite the OPE (\ref{OPE1}) with these variables,
\begin{equation}
 \varphi(Z) \varphi(W) \sim
  - \log (Z - W)^{(r-1)/M} (\omega_M^{n_z}Z^{1/M}-\omega_M^{n_w}W^{1/M}) .
  \label{OPE2}
\end{equation}
We then show the second part of this OPE singularity depends on the
choice of the $M$-th root branch. 
In the case of $n_z=n_w$, its singular behavior is
given by
\begin{eqnarray}
 \log (\omega_M^{n_z}Z^{1/M}-\omega_M^{n_w}W^{1/M}) & \sim &
  \log (Z-W) .
\end{eqnarray}
This is because when we write the right hand side as
$(Z-W)=(Z^{1/M}-W^{1/M})(Z^{(M-1)/M}+Z^{(M-2)/M}W^{1/M}+\cdots+W^{(M-1)/M})$,
the latter part becomes regular at $Z \sim W$.
Thus the OPE (\ref{OPE2}) should be rewritten as
\begin{equation}
  \varphi(Z) \varphi(W) \sim
  - \log (Z-W)^{(r-1)/M+1} .
  \label{OPE3}
\end{equation}
On the other hand, when $n_z\not=n_w$, $\log
(\omega_M^{n_z}Z^{1/M}-\omega_M^{n_w}W^{1/M})$ is regular at $Z \sim W$.
In other words, we cannot obtain
the form of $\log(Z-W)$ from $\log
(\omega_M^{n_z}Z^{1/M}-\omega_M^{n_w}W^{1/M})$, as discussed in the case
of $n_z=n_w$, without modifying its singular behavior at $Z\sim W$ due
to the identity
\begin{equation}
 \omega_M^{(M-1)n} x^{M-1}
  + \omega_M^{(M-2)n} x^{M-2} y
  + \cdots + y^{M-1} = 0
  \quad \mbox{at} \quad x = y , \quad n \not\equiv 0 \ (\mbox{mod}\,M).
\end{equation}
Therefore the OPE (\ref{OPE2}) behaves as
\begin{equation}
   \varphi(Z) \varphi(W) \sim - \log (Z - W)^{(r-1)/M}.
\end{equation}
As a result, the singularity of the OPE (\ref{OPE2}) depends on the choice
of the branch for the $M$-th root of the variable.
Thus we parametrize positions of particles as $z_i = \omega_M^u
(z_I^{(u)})^{1/M}$, satisfying $z_i^M = z_I^{(u)}$.
Then the FQH state, which is represented as the corresponding
conformal block of the primary field, is given by
\begin{equation}
 \Phi(\{z_i^{(u)}\}) 
  = \prod_{u=1}^M \prod_{I<J}^{N^{(u)}} (z_I^{(u)} - z_J^{(u)})^{(r-1)/M+1}
  \prod_{u<v}^M \prod_{I,J} (z_I^{(u)} - z_J^{(v)})^{(r-1)/M} .
  \label{higher_Halperin2}
\end{equation}
This multi-component FQH state is just the $\SU(M)$-singlet Halperin
state (\ref{higher_Halperin}).
Here we obtain this spin-singlet state from only one kind of the $q$-boson field
by taking the root of unity limit.
This is a specific result coming from its $\SU(M)$ symmetry.
We have to introduce the corresponding $M$ types of bosons to
construct a generic $\SU(M)$ state, i.e. a non-singlet state.
Although the singlet-state (\ref{higher_Halperin2}) includes the index
labeling the particle state, $u = 1, \cdots, M$, its role can be freely
changed due to the $\SU(M)$-singlet property.

The filling fraction of this state is obtained by substituting
$r \to (r-1)/M+1$ into the formula (\ref{fraction_M}),
\begin{equation}
 \nu = \frac{M}{M((\frac{r-1}{M}+1)-1)+1} 
  = \frac{M}{r}.
\end{equation}
This is almost the same as the fraction of the Laughlin state $\nu=1/r$,
up to the factor corresponding to the number of internal states.
This factor is interpreted as a result of changing the variables as 
$z_i^M=z_I^{(u)}$.


\section{Spinful FQH states with admissible condition}
\label{sec:admissible}

We extend the relation between the spinless and spin-singlet FQH
wavefunctions to more generalized FQH states.
In this section we describe them in terms of the occupation number
represenatation with emphasis on the admissible condition and its
connection to the Jack polynomial, which is
useful to study the generic FQH states.
The scheme discussed in this section can describe not only the Abelian
FQH state shown in section~\ref{sec:one_to_multi}, but also the
non-Abelian state.
Thus it is expected that the spinless description of the spin-singlet state
is possibe even for the non-Abelian state.

\subsection{Admissible condition}

We introduce the occupation number representation of the FQH states.%
\footnote{See, for example,
\cite{PhysRevLett.100.246802,PhysRevB.77.184502,PhysRevLett.101.246806}.}
We represent a partition $\lambda=(\lambda_1,\lambda_2, \cdots,
\lambda_N)$ with length $\ell_\lambda \le N$ as a (bosonic) occupation
number configuration $n(\lambda)=\{n_m(\lambda), m=0,1,2,\cdots\}$.
This means each of the lowest Landau level (LLL) orbitals,
where $n_m(\lambda)$ is the multiplicity of $m$ in $\lambda$.
We can implement the {\em dominance order} for partitions $\lambda >
\mu$, with the {\em squeezing rule} which connects configurations
$n(\lambda) \to n(\mu)$.
The ground state of the FQH system on the sphere can be uniquely
represented with the {\em root partition}, which is the most dominant
partition with fixing the particle number.

The Jack polynomial is a symmetric polynomial labeled by a
partition $\lambda$, which can be expanded with non-interacting states
(monomial polynomials) obtained by squeezing procedure,
\begin{equation}
 J_\lambda^{\alpha}(\{z_i\}) = m_\lambda(\{z_i\})
  + \sum_{\mu < \lambda} c_{\lambda\mu}(\alpha) m_\mu (\{z_i\}) .
\end{equation}
This is just an eigenstate of the Laplace-Beltrami operator,
\begin{equation}
 \mathcal{H}_{\rm LB} =
 \sum_{i=1}^N \left(z_i \frac{\partial}{\partial z_i}\right)^2
  + \frac{1}{\alpha} \sum_{i<j} \frac{z_i+z_j}{z_i-z_j}
  \left(
   z_i \frac{\partial}{\partial z_i} - z_j \frac{\partial}{\partial z_j} 
  \right) .
  \label{LB_op}
\end{equation}
It has been shown that when the parameter $\alpha$ is negative as $\alpha =
-(k+1)/(r-1)$, the Jack polynomial obeys the admissible condition
\cite{Feigin01012002},
\begin{equation}
 \lambda_{i} - \lambda_{i+k} \ge r .
  \label{ad_cond}
\end{equation}
The root configuration for the $(k,r)$-admissible state, whose $L_z$
component of angular momentum is zero, is given by
$|\Phi_{(k,r)}\rangle=|k0^{r-1}k0^{r-1}k0^{r-1}\cdots\rangle$ and the
corresponding magnetic flux for the spherical system yields
\begin{equation}
 N_\phi = \frac{r}{k} N - r \equiv \frac{1}{\nu} N - r.
  \label{spherical_flux}
\end{equation}
Therefore the filling fraction reads $\nu = k/r$.
The Laughlin state (\ref{Laughlin}) satisfies the $(k=1,r)$
admissible condidtion, $|\Phi_{\rm L} \rangle = |10^{r-1}10^{r-1} \cdots
\rangle$.
Note that the non-Abelian FQH states, e.g. Moore-Read
\cite{Moore:1991ks} and Read-Rezayi states \cite{PhysRevB.59.8084}, can be
also described in this way: the RR state associated with $\Z_k$
parafermion obeys the $(k,2)$-admissible condition
\cite{PhysRevLett.100.246802,PhysRevB.77.184502,PhysRevLett.101.246806}.

We then consider the mapping from the spinless to the
spin-singlet states, discussed in section \ref{sec:one_to_multi}, in
terms of these representations.
The corresponding partition to the root configuration for the $(k,r,M=1)$
state $|k0^{r-1}k0^{r-1}\cdots\rangle$ is given by
\begin{eqnarray}
  \lambda & = & (\lambda_1, \lambda_2, \cdots, \lambda_N)
   \nonumber \\
 & = &
  (
   \underbrace{
   r \left(N/k-1\right), \cdots, 
   r \left(N/k-1\right)}_k , 
   \cdots, 
   \underbrace{ 2r, \cdots, 2r }_k,
   \underbrace{ r, \cdots, r }_k,
   \underbrace{ 0, \cdots, 0 }_k   
  ) 
  \label{root_config}
\end{eqnarray}
where $N$ is the particle number and we can see $\lambda_1 = N_\phi$
defined in (\ref{spherical_flux}).
In order to obtain the spin-singlet state as discussed in
section~\ref{sec:one_to_multi}, we then introduce the following
partition from this root configuration (\ref{root_config}), which
corresponds to the modified variables (\ref{change_variable}),
\begin{eqnarray}
  \tilde \lambda & = &
  \left(
   \left[\frac{\lambda_1}{M}\right], \cdots,
   \left[\frac{\lambda_{N-1}}{M}\right], 
   \left[\frac{\lambda_N}{M}\right]
  \right)
  \nonumber \\
 & = & 
  (
   \underbrace{
   [r \left(N/k-1\right)/M], \cdots, 
   [r \left(N/k-1\right)/M]}_k , 
   \cdots, 
   \underbrace{ [r/M], \cdots, [r/M] }_k,
   \underbrace{ [0], \cdots, [0] }_k   
  ) .
  \nonumber \\
\end{eqnarray}
Here $[x]$ denotes the floor function, providing the largest integer not
greater than $x$.
We can read the following relation between the parameters of the
spinless and the singlet state from the expression shown in
(\ref{higher_Halperin2})
\begin{equation}
 \tilde r = \frac{r-1}{M} + 1 
  \qquad \Longleftrightarrow \qquad
  r = M(\tilde r - 1) + 1.
 \label{filling_correspondence}
\end{equation}
Thus each component is written as
\begin{equation}
 \frac{r}{M} p
 = p (\tilde r - 1) + \frac{p}{M}, \qquad p \in \Z .
\end{equation}
Therefore we have
\begin{equation}
 \left[
  \frac{r}{M} p
 \right]
 -  
 \left[
  \frac{r}{M} (p-1)
 \right]
 = \left\{
    \begin{array}{cccc}
      \tilde r - 1 & \mbox{for} & p \not\equiv 0 &
      (\mbox{mod}~ M) \\
     \tilde r & \mbox{for} & p \equiv 0 &
      (\mbox{mod}~ M) \\
    \end{array}
   \right. .
\end{equation}
In the occupation number basis this configuration is represented as
\begin{equation}
 |X0^{\tilde r-1} X 0^{\tilde r-1} \cdots \rangle
  \qquad \mbox{with} \quad
  X=k^{(0)}0^{\tilde r-2} k^{(1)} 0^{\tilde r-2} \cdots k^{(M-1)} .
  \label{root_config_M}
\end{equation} 
Superscripts stand for $\lambda_i$ modulo $M$, and correspond to
internal degrees of freedom of the $\SU(M)$-singlet state.
This is just the root configuration for the sipn-singlet FQH state
\cite{Ardonne:2011,Estienne:2011}.
Note that the state $u=0$ is equivalent to $u=M$.

The $\SU(M)$-singlet $(k,r)$-state is obtained from the spin
Laplace-Beltrami operator \cite{{Estienne:2011}}
\begin{equation}
  \mathcal{H}_{\rm sLB} =
 \sum_{i=1}^N \left(z_i \frac{\partial}{\partial z_i}\right)^2
  + \frac{1}{\alpha} \sum_{i<j} \frac{z_i+z_j}{z_i-z_j}
  \left(
   z_i \frac{\partial}{\partial z_i} - z_j \frac{\partial}{\partial z_j} 
  \right) 
  - \frac{1}{\alpha} \sum_{i\not=j} (1-K_{ij}) \frac{z_i z_j}{(z_i-z_j)^2},
  \label{sLB_op}
\end{equation}
where $K_{ij}$ stands for the exchange operator.
In this case the condition (\ref{ad_cond}) is modified as
\begin{equation}
 \lambda_i - \lambda_{i+k} \ge r - 1.
\end{equation}
If $\lambda_i - \lambda_{i+k} = r - 1$, the spin part satisfies
\begin{equation}
 \sigma_i > \sigma_{i+k}.
\end{equation}
Note that the definition we use in this paper is slightly different from the
notation used in \cite{{Estienne:2011}}.
Thus we can see (\ref{root_config_M}) is just the root configuration for
the $\SU(M)$-singlet $(k,\tilde r)$-state.

Conversely, starting with an $M$-component state labeled by
$(\tilde\lambda,\sigma)$, where the latter stands for the
internal degree with $0\le \sigma \le M-1$, the corresponding
spinless state can be recovered as
\begin{equation}
 \lambda_i = M \tilde \lambda_i + \sigma_i .
\end{equation}
Essentially this relation has been already shown in
\cite{Uglov:1997ia,Dijkgraaf:2007fe,Kimura:2011zf,Kimura:2011gq}, and
corresponds to the method discussed in section~\ref{sec:one_to_multi}.
This translation procedure is already explicitly shown in
\cite{Uglov:1997ia} for the positive parameter case $\alpha>0$.

\subsection{Squeezing rule}

\begin{figure}[t]
 \begin{center}
  \vspace{-5em}
  \includegraphics[width=35em]{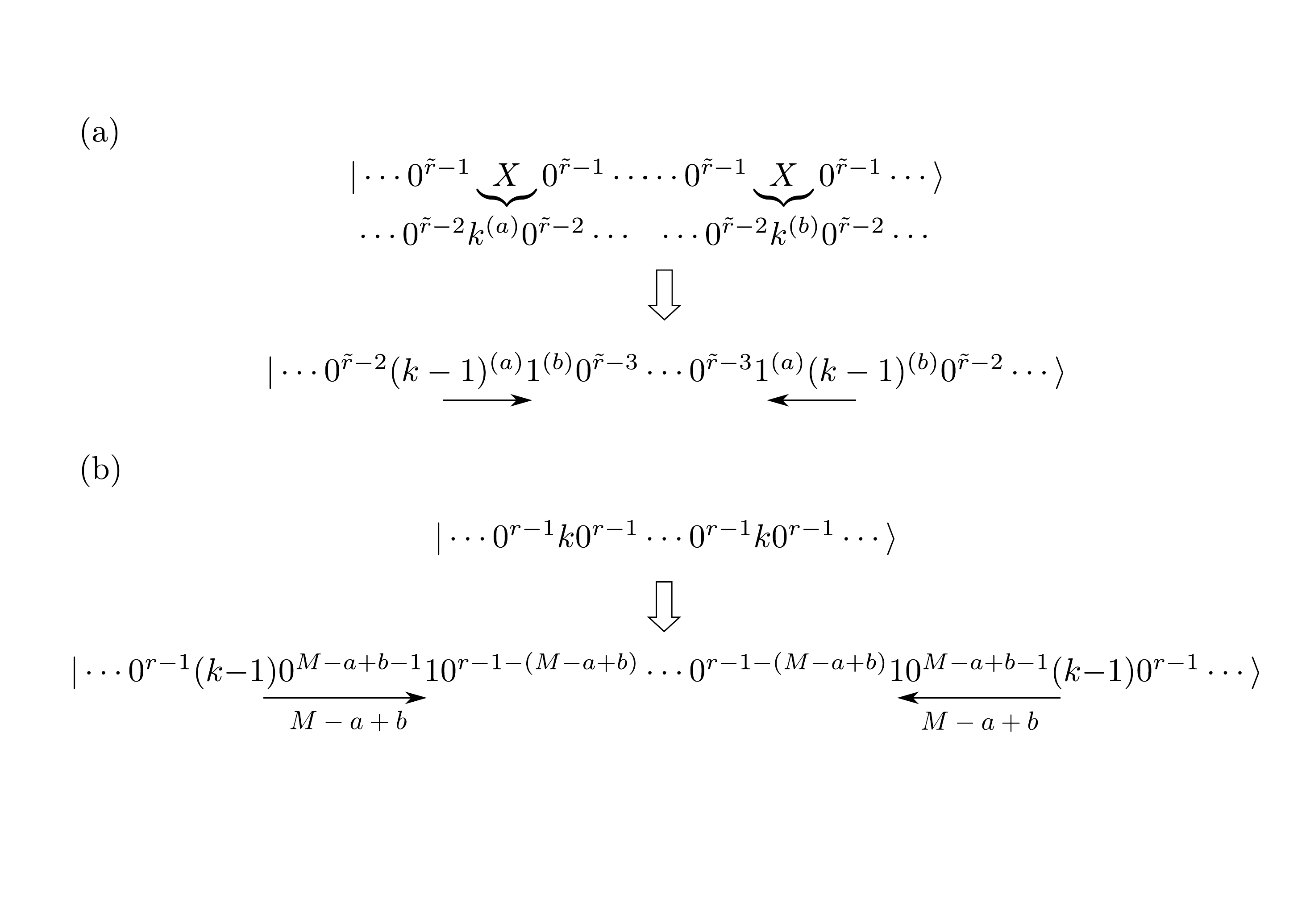}
  \vspace{-6em}
 \end{center}
 \caption{The squeezing rule for the singlet admissible states in terms
 of (a) the original spinful configuration and (b) the spinless
 description. The parameters are related as $r = M(\tilde r - 1)+1$.}
 \label{sq_fig}
\end{figure}

We study the squeezing rule for the spin-singlet state in
terms of the spinless basis.
From the root configuration we squeeze $a$-th and $b$-th components in
$i$-th and $j$-th blocks of $X$ given in (\ref{root_config_M}), and
exchange their states, as shown in Fig.~\ref{sq_fig}.
First we consider the case $i<j$.
This operation is represented in terms of $(\tilde \lambda,\sigma)$ as
\begin{eqnarray}
 \left((i-1)(M(\tilde r-1)+1) + a(\tilde r-1) , a \right)
  & \longrightarrow &
  \left((i-1)(M(\tilde r-1)+1) + a(\tilde r-1) + 1 , b \right) ,
  \nonumber \\
 \left((j-1)(M(\tilde r-1)+1) + b(\tilde r-1) , b \right)
  & \longrightarrow &
  \left((j-1)(M(\tilde r-1)+1) + b(\tilde r-1) - 1 , a \right) .
  \nonumber \\
\end{eqnarray}
The corresponding operation in terms of the spinless basis $\lambda$ is
just given by squeezing $M-a+b$ boxes,
\begin{eqnarray}
 M[(i-1)(M(\tilde r-1)+1)+a(\tilde r-1)] + a
  & \longrightarrow &
 M[(i-1)(M(\tilde r-1)+1)+a(\tilde r-1)+1] + b ,
 \nonumber \\
 M[(j-1)(M(\tilde r-1)+1)+b(\tilde r-1)] + b
  & \longrightarrow &
 M[(j-1)(M(\tilde r-1)+1)+b(\tilde r-1)-1] + a .
 \nonumber \\
\end{eqnarray}
We can see $M-a+b>0$, since they satisfy $0 \le a,b \le M-1$.

Next let us study the case $i=j$.
This operation is given by
\begin{eqnarray}
 \left((i-1)(M(\tilde r-1)+1) + a(\tilde r-1) , a \right)
  & \longrightarrow &
  \left((i-1)(M(\tilde r-1)+1) + a(\tilde r-1) , b \right) ,
  \nonumber \\
 \left((j-1)(M(\tilde r-1)+1) + b(\tilde r-1) , b \right)
  & \longrightarrow &
  \left((j-1)(M(\tilde r-1)+1) + b(\tilde r-1) , a \right) ,
  \nonumber \\
\end{eqnarray}
and the corresponding one yields
\begin{eqnarray}
 M[(i-1)(M(\tilde r-1)+1)+a(\tilde r-1)] + a
  & \longrightarrow &
 M[(i-1)(M(\tilde r-1)+1)+a(\tilde r-1)] + b ,
 \nonumber \\
 M[(j-1)(M(\tilde r-1)+1)+b(\tilde r-1)] + b
  & \longrightarrow &
 M[(j-1)(M(\tilde r-1)+1)+b(\tilde r-1)] + a .
 \nonumber \\
\end{eqnarray}
In this case the number of squeezed boxes is $b-a$.
Therefore we can perform this operation only when $b>a$.
Fig.~\ref{sq_fig} shows the squeezing rule for the $\SU(M)$ states
and the corresponding spinless description.

As an example we now consider the $(k,\tilde r,M)=(1,2,2)$ state,
which is well investigated in \cite{Estienne:2011}.
The corresponding spinless state yields the $(k,r,M)=(1,3,1)$ state.
The root configuration for $N=4$ is given by
\begin{equation}
 |\uparrow, \downarrow, 0,  \uparrow, \downarrow \rangle .
  \label{root02}
\end{equation}
This is also represented by the following partitions,
\begin{equation}
 (\tilde \lambda,\sigma) = ( 4^{(\uparrow)}, 3^{(\downarrow)}, 1^{(\uparrow)}, 0^{(\downarrow)})
  \quad \Longleftrightarrow \quad
  \lambda =(9,6,3,0),
\end{equation}
where we assign $\uparrow \, \equiv 1$, $\downarrow \, \equiv 0$ (mod 2).
The descendants of the root configuration (\ref{root02}) are given by
\begin{eqnarray}
 \begin{array}{ccccc}
  |\uparrow, \downarrow, 0, \downarrow, \uparrow \rangle & = &
  (4^{(\uparrow)}, 3^{(\downarrow)}, 1^{(\downarrow)}, 0^{(\uparrow)}) & \Longleftrightarrow &
  (9,6,2,1)  \\
  |\downarrow, \uparrow, 0, \uparrow, \downarrow \rangle & = &
  (4^{(\downarrow)}, 3^{(\uparrow)}, 1^{(\uparrow)}, 0^{(\downarrow)}) & \Longleftrightarrow &
  (8,7,3,0)  \\
  |\uparrow, \downarrow, 0, \uparrow, \downarrow \rangle & = &
  (4^{(\downarrow)}, 3^{(\uparrow)}, 1^{(\downarrow)}, 0^{(\uparrow)}) & \Longleftrightarrow &
  (8,7,2,1)  \\
  |\uparrow, \uparrow, 0, \downarrow, \downarrow \rangle & = &
  (4^{(\downarrow)}, 3^{(\downarrow)}, 1^{(\uparrow)}, 0^{(\uparrow)}) & \Longleftrightarrow &
  (8,6,3,1)  \\
 \end{array}
\end{eqnarray}
The last one is equivalent to $|\downarrow,\downarrow, 0, \uparrow,
\uparrow \rangle$ due to the spin-singlet condition, which is also
represented as
\begin{equation}
 |\downarrow,\downarrow, 0, \uparrow,\uparrow \rangle
  \quad = \quad
 (4^{(\uparrow)},3^{(\uparrow)},1^{(\downarrow)},0^{(\downarrow)})
  \quad \Longleftrightarrow \quad
 (9,7,2,0) .
\end{equation}
Note that $(9,7,2,0)$ cannot be obtained from the root configuration
$(9,6,3,0)$ by squeezing it.
Therefore, to be consistent with the symmetry,
it turns out to be the zero-weight state, as discussed in
\cite{Ardonne:2011,Estienne:2011}.

\section{Relation to conformal field theory}
\label{sec:CFT}

One of the most important properties of the FQH state is its exotic
statistics.
Such a crucial property can be well described in terms
of the two dimensional CFT: the non-Abelian
statistics is interpreted as a consequence of the fusion rule of the
corresponding CFT.
In this paper we have discussed a novel description of the spin-singlet
state by utilizing the $q$-deformed state.
Thus, in this section, we comment on CFT relevant to the spin-singlet FQH
state with emphasis on its relation to the $q$-deformed CFT.

The typical underlying CFT for the FQH state is the $\Z_k$-parafermion model.
It describes the $k$-th Read-Rezayi state \cite{PhysRevB.59.8084}, which
includes the Moore-Read state \cite{Moore:1991ks} at $k=2$.
The $\Z_k$-parafermion model can be obtained from the coset
construction,
\begin{equation}
 \frac{\SU(2)_k}{\U(1)}.
  \label{Zk_parafermion}
\end{equation}
Furthermore, it has been shown the generic $(k,r)$-admissible state is
coming from the extended chiral algebra, $\mathrm{WA}_{k-1}(k+1,k+r)$
\cite{Bernevig:2009JPhA,1751-8121-42-44-445209,PhysRevB.82.205307}.
A CFT model for the FQH state has been also investigated for the
spin-singlet states.
The $\SU(M)$-singlet state corresponds to the generalized parafermion
model,
\begin{equation}
 \frac{\SU(M+1)_k}{\U(1)^M}.
\end{equation}
This is a natural extention of the $k$-th Read-Rezayi state
\cite{PhysRevLett.82.5096,Ardonne:2001NuPhB}.

The interesting property of the parafermion models is the level-rank
duality.\footnote{See a review article, e.g. \cite{Bouwknegt:1992wg}.}
For example, for the $k$-th Read-Rezayi state, we have
\begin{equation}
 \frac{\SU(2)_k}{\U(1)} = \frac{\SU(k)_1\times\SU(k)_1}{\SU(k)_2}.
\end{equation}
This means the $\Z_k$-parafermion is also realized as the lowest one
of the $\mathcal{W}$-minimal series,
\begin{equation}
 \frac{\SU(k)_l \times \SU(k)_1}{\SU(k)_{l+1}},
  \label{W_minimal}
\end{equation}
which gives rise to the central charge,
\begin{equation}
 c = k - 1 - k (k^2-1) \frac{(p-q)^2}{pq}.
\end{equation}
Here we identify $k+l = p/(q-p)$, $k+l+1 = q/(q-p)$.
This duality is also applied to the spin-singlet theory,
\begin{equation}
 \frac{\SU(M+1)_k}{\U(1)^M} 
  = \frac{(\SU(k)_1)^{M+1}}{\SU(k)_{M+1}}.
  \label{dual_M_comp}
\end{equation}
Recently the corresponding CFT series, which reproduces the model in the
right hand side of (\ref{dual_M_comp}) at the lowest level, has been
proposed \cite{Estienne:2011},
\begin{equation}
 \frac{\SU(k)_l\times(\SU(k)_1)^M}{\SU(k)_{M+l}}.
  \label{W_minimal2}
\end{equation}
Substituting $k+l=Mp/(q-p)$, $k+l+M=Mq/(q-p)$, the central
charge is given by
\begin{equation}
 c = M(k-1) - \frac{k(k^2-1)}{M} \frac{(p-q)^2}{pq}.
\end{equation}
This novel CFT series is not well investigated yet, but let us comment
on its connection to the four dimensional gauge theory.
Recent progress on the supersymmetric gauge theory reveals the
remarkable relation between the four dimensional gauge theory and the
two dimensional CFT \cite{Alday:2009aq}: the instanton partition
function of the gauge theory is directly interpreted as the conformal
block of the two dimensional CFT.%
\footnote{An attempt to connect the gauge theory partition function with
the FQH state is found in \cite{Santachiara:2010bt}.}
The standard $\SU(N)$ Yang-Mills theory corresponds to the 
$A_{N-1}$ Toda CFT.
Its central charge is given by
\begin{equation}
 c = N - 1 + N(N^2-1) Q^2,
\end{equation}
where $Q$ is related to the regularization parameter of the four
dimensional theory.
This CFT is essentially the same as the model shown in
(\ref{W_minimal}): they are equivalent under the identification
$Q^2=-(p-q)^2/(pq)$.

A similar connection is also suggested for the generalized model
(\ref{W_minimal2}).
In this case the gauge theory on the type $A_{M-1}$ ALE space, which is
given by resolving the singularity of the orbifold $\C^2/\Z_M$, gives
rise to the corresponding CFT \cite{Nishioka:2011jk}.
Its central charge is given by
\begin{equation}
 c = M(N-1) + \frac{N(N^2-1)}{M} Q^2 .
\end{equation}
When we consider a gauge group $G$, the central charge is written
in a generic form,
\begin{equation}
 c = M r_G + \frac{d_G h_G}{M} Q^2,
  \label{generic_model_center}
\end{equation}
where $r_G$, $d_G$ and $h_G$ are rank, dimension and the dual
Coxeter number of the group $G$, respectively.
This generalized CFT would be realized as the following coset model,
\begin{equation}
 \frac{G_l \times (G_1)^M}{G_{M+l}}.
\end{equation}

Let us then comment on the $q$-deformed CFT.
It has been shown that, by taking the limit, $t=q^r$ and then $q\to 1$,
the central charge of the corresponding CFT is given by
\cite{Shiraishi:1995rp,Awata:1995zk}
\begin{equation}
 c = 1 - 6 \frac{(1-r)^2}{r}.
\end{equation}
When we start with the $q$-deformed $\mathcal{W}$ CFT, it becomes
\begin{equation}
 c = N-1 - N(N^2-1) \frac{(1-r)^2}{r}.
\end{equation}
Indeed they are equivalent to the $\SU(N)$ minimal models via $r=p/q$.
Thus, when we apply the root of unity limit, $q\to\omega_M q$, $t\to \omega_M
q^r$ and $q\to 1$, it is natural to obtain the CFT, which describes the
$\SU(M)$-singlet FQH state as a result of the Yangian Gelfand-Zetlin basis.
Its central charge is expected to be given by
\begin{equation}
 c = M(N-1) - \frac{N(N^2-1)}{M} \frac{(1-r)^2}{r}.
\end{equation}
This model corresponds to the standard Macdonald polynomial, which is
associated with the type $A$ root system.
Thus we now expect that the central charge of the model, which is coming
from the $q$-deformed theory related to other root systems, e.g. the $BC$
type theory \cite{Koomwinder:1992,Kasatani:2003}, can be written in a
form similar to (\ref{generic_model_center}).

\section{Conclusion}
\label{sec:summary}

We have investigated the $\SU(M)$-singlet FQH states with the spinless basis.
We have shown the raising and lowering operators for $\SU(M)$ states
with emphasis on its similarity to the standard $\SU(2)$ spinful states.
The $\SU(M)$-singlet condition can be written in a quite similar
form to the $\SU(2)$-singlet condition for the Halperin state.
We have obtained the $\SU(M)$-singlet Halperin state from the
corresponding Laughlin state by considering the $q$-deformation and its
root of unity limit.
This is just the prescription to impliment the Yangian Gelfand-Zetlin
basis in terms of a certain spinless state, which is proposed in
\cite{Uglov:1997ia}.
As well known, the FQH trial wavefunction is regarded as the correlation
function of the primary fields in the corresponding CFT.
Thus, to discuss such a correlation function, we have studied the
$q$-boson fields and its OPE in the root of unity limit.
We have also shown the relation between the $\SU(M)$-singlet and the
corresponding spinless states for much generalized FQH states,
which obey the $(k,r)$-admissible condition.
We have discussed them in terms of the occupation number representation of the
FQH states, and thus the squeezing rule is naturally assigned for them.
We then have commented on the underlying CFT for the FQH states
discussed in this paper.
There would be the interesting structure in the CFT for the
multi-component states, and the root of unity limit of the $q$-deformed CFT.

\subsection*{Acknowledgments}

The author would like to thank Y.~Hidaka and T.~Nishioka for useful comments.
The author is supported by Grant-in-Aid for JSPS Fellows (No.~23-593).


\bibliographystyle{ytphys}

\bibliography{/Users/k_tar/Dropbox/etc/conf}

\end{document}